\documentclass[12pt]{article}
\usepackage{amssymb,amsmath}
\usepackage{amsthm}
\usepackage{graphicx}
\usepackage{enumerate}
\usepackage{cite}
\usepackage{url} 
\usepackage{colortbl}
\usepackage{color}
\usepackage{authblk}

\title{\bf Fitting the Distribution of Linear Combinations of $t-$Variables with more than 2 Degrees of Freedom}
\author[1]{Onel L. A. L\'opez}
\author[2]{Evelio G. Fern\'andez}
\author[1]{Matti Latva-aho}

\affil[1]{Centre for Wireless Communications, University of Oulu, Finland, onel.alcarazlopez@oulu.fi, matti.latva-aho@oulu.fi}
\affil[2]{Department of Electrical Engineering, Federal 	University of Parana, Brazil, evelio@ufpr.br}

\date{}

\begin{document}
	
	\maketitle
	
	\def\spacingset#1{\renewcommand{\baselinestretch}%
		{#1}\small\normalsize} \spacingset{1}

\begin{abstract}
The linear combination of Student's $t$ random variables (RVs) appears in many statistical applications. Unfortunately, the Student's $t$ distribution is not closed under convolution, thus, deriving an exact and general distribution for the linear combination of $K$ Student's $t$ RVs is infeasible, which motivates a fitting/approximation approach. Here, we focus on the scenario where the only constraint is that the number of degrees of freedom of each $t-$RV is greater than two. Notice that since the odd moments/cumulants of the Student's $t$ distribution are zero, and the even moments/cumulants do not exist when their order is greater than the number of degrees of freedom, it becomes impossible to use conventional approaches based on moments/cumulants of order one or higher than two. To circumvent this issue, herein we propose fitting such a distribution to that of a scaled Student's $t$ RV by exploiting the second moment together with either the first absolute moment or the characteristic function (CF). 
For the fitting based on the absolute moment, we depart from the case of the linear combination of $K= 2$ Student's $t$ RVs and then generalize to $K\ge 2$ through a simple iterative procedure. Meanwhile, the CF-based fitting is direct, but its accuracy (measured in terms of the Bhattacharyya distance metric) depends on the CF parameter configuration, for which we propose a simple but accurate approach.
We numerically show that the CF-based fitting usually outperforms the absolute moment -based fitting and that both the scale and number of degrees of freedom of the fitting distribution increase almost linearly with $K$.
\end{abstract}

\vfill

\newpage
\spacingset{1.9} 
\section{Introduction}
\label{sec:intro}

The Student's $t$-distribution arises in numerous scenarios, e.g., when estimating the mean of a normally distributed population of unknown   variance with relatively few samples, and in Bayesian analysis of data from a normal family. 
Moreover, such a distribution plays a key role in many relevant statistical analyses, including Student's $t$-test for assessing the statistical significance of the difference between two sample means, the construction of confidence intervals for the difference between two population means, and in linear regression analysis \cite{Bening.2005,Glenn.1978,Witkovsk.2002,Berg.2008,Nadarajah.2005,Mohammad.2014}. 

One of the distinctive properties of the Student's $t$ distribution is its heavy tail. This behavior is also seen in  the famous family of stable distributions, however, the Student's $t$ distribution is more analytically tractable, which allows, for example, to write down explicitly its likelihood function \cite{Bening.2005}.

\subsection{Main Statistics of the Student's $t$ Distribution}
The probability density function (PDF) and cumulative density function (CDF) of a Student's $t$ random variable (RV) $T$ with $\nu\in\mathbb{R}^+$ degrees of freedom, i.e., $T\sim\mathcal{T}(\nu)$, is given by \cite{Mohammad.2014}
\begin{align}	
	f_T(x)&=\alpha_\nu\Big(1+\frac{x^2}{\nu}\Big)^{-\frac{\nu+1}{2}},\label{fT}\\
	F_T(x)&= 1-\frac{1}{2}\mathcal{I}_{\frac{\nu}{x^2+\nu}}\Big(\frac{\nu}{2},\frac{1}{2}\Big),\label{FT}
\end{align}
$\forall x\in\mathbb{R}$, 
where $\mathcal{I}_x(\cdot,\cdot)$ is  regularized incomplete beta function \cite[eq. (8.17.2)]{Olver.2010}, and
\begin{align}
	\alpha_\nu\triangleq 	\frac{1}{\sqrt{\nu}B\big(\frac{\nu}{2},\frac{1}{2}\big)}= \frac{\Gamma((\nu+1)/2)}{\Gamma(\nu/2)\sqrt{\nu\pi}}.\label{anu}
\end{align}
Here, $B(\cdot,\cdot)$ and $\Gamma(\cdot)$ are the beta \cite[eq. (5.12.1)]{Olver.2010} and gamma \cite[eq. (5.2.1)]{Olver.2010} functions, respectively. Meanwhile, the integer moments are \cite{Mohammad.2014}
\begin{align}
	\mathbb{E}[T^m]=\nu^{m/2}\prod_{i=1}^{m/2}\frac{2i-1}{\nu-2i},\ m\ \text{even},\ m<\nu,\label{Er}
\end{align}
while $\mathbb{E}[T^m]=0$ for $m$ odd, and moments of order $\nu$ or higher do not exist. Observe that for the specific case of $m=2$ (second moment), \eqref{Er} reduces to $\nu/(\nu-2)$.

Two statistics that play  a key role in our proposed approach are the absolute moments and the characteristic function (CF). The latter is given by \cite{Mohammad.2014}
\begin{align}
	\mathrm{CF}_T(r) &=\mathbb{E}[e^{\dot{\iota} |r|T}]= \frac{(\sqrt{\nu}|r|)^{\nu/2}K_{\nu/2}(\sqrt{\nu}|r|)}{2^{\nu/2-1}\Gamma(\nu/2)},\ \nu >0,\label{CF}
\end{align}
where $K_{\nu}(\cdot)$ is the modified Bessel function of second kind and order $\nu$ \cite[Sec. 10.25]{Olver.2010}.\footnote{Notice that $\mathrm{CF}_T(r)\in\mathbb{R}$, while the moment generating function of $T$ does not exist.}  Meanwhile, the absolute moments can be obtained as follows
\begin{align}
	\mathbb{E}\big[|T|^m\big]&=\int_{-\infty}^{\infty}|x|^m f_T(x)\mathrm{d}x\nonumber\\
	&\stackrel{(a)}{=}2\alpha_\nu\int_{0}^{\infty}x^m\Big(1+\frac{x^2}{\nu}\Big)^{-(\nu+1)/2}\mathrm{d}x\nonumber\\ 
	&\stackrel{(b)}{=}\frac{\alpha_\nu \nu^{\frac{m+1}{2}}\Gamma\big(\frac{\nu-m}{2}\big)\Gamma\big(\frac{m+1}{2}\big)}{\Gamma((\nu+1)/2)}\ (\text{for } m<\nu)\nonumber\\
	&\stackrel{(c)}{=}\frac{ \nu^{m/2}\Gamma\big(\frac{\nu-m}{2}\big)\Gamma\big(\frac{m+1}{2}\big)}{\sqrt{\pi}\Gamma(\nu/2)}, \ m<\nu,\label{abs}
\end{align}
where (a) follows from leveraging the symmetry of $T$ around zero and from substituting \eqref{fT}, (b) exploits \cite[eq. (3.241.4)]{Gradshteyn.2014} to solve the definite integral, and (c) is attained after simple algebraic transformations after substituting \eqref{anu}.
\subsection{On the linear combination of Student's $t$ RVs}\label{LCS}
The linear combination of Student's $t$ RVs, denoted as
\begin{align}
	Z\triangleq \sum_{i=1}^K \sigma_iT_i, \label{Zi}
\end{align}
where $T_i\sim \mathcal{T}(\nu_i)$ and (without loss of generality) $\sigma_i>0$,  appears in many statistical applications. For instance, 
\begin{itemize}
	\item Fairweather \cite{Fairweather.1972} proposed a method based on the pivotal quantity $Z$ to obtain an accurate confidence interval for the common mean of several normal populations. Notice that the problem of characterizing the distribution of independent samples that are collected from different normal populations with a common mean but possibly with different variances appear in many practical application, e.g., when different instruments/methods/laboratories are used to measure substances or products to assess their average quality \cite{Krishnamoorthy.2003};
	\item The Behrens-Fisher distribution of the test statistic for testing the equality of the means of two normal populations with unknown variances is that of a linear combination of two independent Student's $t$ RVs. The problem appears in many traditional statistical problems, e.g., check \cite{Kim.1998,Witkovsk.2002,Weerahandi.2003,Hollander.2013};
	\item The distribution of RVs $T_i$ can approximate other heavy-tailed symmetric distributions, e.g., $ X/Y\ \!|\!\ Y\ge y_0$, where $X$ and $Y$ are respectively Gaussian and Rayleigh-distributed. In such scenarios, the distribution of their linear combination $Z$ may be extremely valuable. Interestingly, the sum of random variables of the form $ X/Y\ \!|\!\ Y\ge y_0$ appears in the scenario proposed in \cite{Lopez.2022}, where the goal is to determine the number of active devices in a machine-type wireless communication network by relying on coordinated pilot transmissions  without much signaling overhead, which facilitates the posterior data decoding procedures.
\end{itemize}

Unfortunately, the Student's $t$ distribution is not closed under convolution \cite{Petroni.2007}, thus, deriving the exact distribution of $Z$ has been shown to be a cumbersome task, especially for an arbitrary number of degrees of freedom $\nu_i$ and  number of addends $K$. For instance, the methods proposed in \cite{Petroni.2007,Glenn.1978,Witkovsk.2002,Nadarajah.2005,Berg.2008} are restricted to the case of all $T_i$s having an odd number of degrees of freedom. Meanwhile, the PDF of $Z$ is given in  \cite{Berg.2010} as an infinite series but only for the specific case of $K=2$.

In general, approximation methods are often more tractable and appealing, which motivates our work in this paper. Specifically, we aim to accurately approximate  the distribution of $Z$ in closed-form given that $\{T_i\}$ are independently distributed with $\nu_i> 2$, $\forall i$, and no other constraints.\footnote{The assumption that RVs $\{T_i\}$ are independent is common in the literature, e.g., \cite{Glenn.1978,Berg.2008,Nadarajah.2005,Mohammad.2014,Berg.2010}.} To the best of our knowledge, this is the first work to (satisfactorily) address this.
\subsection{Our approach}\label{our}
For the approximation, we resort to a Student's $t$ distribution fitting. Specifically, we aim to accurately fit $Z\sim \sigma_z\mathcal{T}(\nu_z)$ with $\nu_z>2$, which should hold, at least intuitively, as both distributions share the same symmetric and bell-shaped form. However, what might appear to be a simple and straightforward approach is not when considering that the Student's $t$ distributions are with more than two degrees of freedom and no other constraints. We elaborate this as follows.
	
The distribution fitting approaches commonly rely on moments (including $L-$moments \cite{Headrick.2011}) or cumulants matching. Specifically, at least two moments and/or cumulants of $Z$ are needed to match those of a scaled Student's $t$ distribution since such a distribution is characterized only by the scale $\sigma_z$, and the number of degrees of freedom $\nu_z$. However, the challenge lies in that $\nu_z$ (and each $\nu_i$) must be greater than the moment/cumulant order, while i) the odd moments/cumulants cannot be used since they are zero, and ii) the negative moments do not converge since $f_Z(0)>0$. This implies that i) we cannot fit moments/cumulants of order higher than 2 in order to allow $\nu_z\in(2,\infty)$, and ii) we cannot rely on the first moment/cumulant. Meanwhile, fractional moments could be used, but they are complex and difficult to compute in general. 

In this work, we resort to a fitting based on the second moment matching together with absolute moment or CF matching to circumvent the above issues. Note that using second moment matching is a natural choice given its simplicity, while exploiting the absolute moment seems also appealing. However, although absolute moments of any order, $\mathbb{E}[|\cdot|^{m}]$, with $m\in\mathbb{R}, m<\nu$, could be used, they are cumbersome to derive due to the limited separability of the absolute value of a sum, thus, we focus on the simplest $m=1$ case. Finally, a CF matching performance is intriguing as it is not a commonly adopted approach in the literature for distribution fitting problems, specially because moments or other simpler statistics are often available, so we adopt it here given the special characteristics/challenges of the considered problem.

\section{Computation of the Relevant Statistics of $Z$}

\subsection{Second Moment}\label{sm}
Therefore, the second moment of $Z$ is given by
\begin{align}
	\mathbb{E}[Z^2]=\sum_{i=1}^{K}\mathbb{E}[T_i^2]=\sum_{i=1}^{K}\frac{\sigma_i^2\nu_i}{\nu_i-2},\label{Z1}
\end{align}
which comes from leveraging the independence and zero-mean features of $\{T_i\}$ and from using \eqref{Er} with $m=2$.

\subsection{Characteristic Function}
The CF of the sum of independent RVs matches the product of their independent CFs, thus,
\begin{align}
	\mathrm{CF}_Z(r)&=\prod_{i=1}^{K}\mathrm{CF}_{\sigma_i T_i}(r)=\prod_{i=1}^{K}\mathrm{CF}_{ T_i}(\sigma_i r)\nonumber\\
	&=\prod_{i=1}^{K}\frac{(\sqrt{\nu_i}\sigma_i|r|)^{\nu_i/2}K_{\nu_i/2}(\sqrt{\nu_i}\sigma_i|r|)}{2^{\nu_i/2-1}\Gamma(\nu_i/2)},\ \forall \nu_i >0\nonumber\\
	&=2^{K}(|r|/2)^{\sum_{i=1}^{K}\nu_i/2}\prod_{i=1}^{K}\frac{(\sqrt{\nu_i}\sigma_i)^{\nu_i/2}K_{\nu_i/2}(\sqrt{\nu_i}\sigma_i|r|)}{\Gamma(\nu_i/2)},\ \forall \nu_i >0.\label{CF2}
\end{align}
\subsection{Absolute Moment}
The absolute moment of $Z$ obeys
\begin{align}
	\mathbb{E}[|Z|]=\mathbb{E}\bigg[\Big|\sum_{i=1}^{K}\sigma_iT_i\Big|\bigg].\label{Z2}
\end{align}
Remarkably, further simplifying \eqref{Z2} is not a trivial task. Furthermore, its computation complexity scales with $K$. Therefore, we focus on the case $K=2$, but leverage the corresponding results for the distribution fitting of the linear combination of any $K\ge 2$ RVs in Section~\ref{fitting}.

The absolute moment for the case of $K=2$ can be computed as follows
\begin{align}
	\mathbb{E}[|Z|]&=\mathbb{E}\big[|\sigma_1T_1+\sigma_2T_2|\big]\nonumber\\
	&\stackrel{(a)}{=}2\int\limits_{-\infty}^{\infty}\int\limits_{-\frac{\sigma_2x_2}{\sigma_1}}^{\infty}(\sigma_1x_1+\sigma_2x_2)f_{T_1}(x_1)f_{T_2}(x_2)\mathrm{d}x_1\mathrm{d}x_2\nonumber\\
	&\stackrel{(b)}{=}2\sigma_1\underbrace{\int\limits_{-\infty}^{\infty}\int\limits_{\frac{\sigma_2|x_2|}{\sigma_1}}^{\infty}x_1f_{T_1}(x_1)f_{T_2}(x_2)\mathrm{d}x_1\mathrm{d}x_2}_{I_1}+2\sigma_2\underbrace{\int\limits_{-\infty}^{\infty}\int\limits_{-\frac{\sigma_2x_2}{\sigma_1}}^{\infty}x_2f_{T_1}(x_1)f_{T_2}(x_2)\mathrm{d}x_1\mathrm{d}x_2}_{I_2}, \label{EZ}
\end{align}
where in (a) we exploit the fact that $\sigma_1T_1+\sigma_2T_2$ is symmetric around 0, thus, it can adopt positive and negative values with probability 0.5. Then, (b) comes after applying the integral operator to each of the integrand's addends and leveraging the symmetry of $\int_{-a}^ax_1 f_{T_1}(x_1)\mathrm{d}x_1=0$. Now, observe that $\mathbb{E}[|Z|]=2(\sigma_1 I_1+\sigma_2 I_2)$, and we are concerned with computing $I_1$ and $I_2$. 

In the case of $I_1$, we have that
\begin{align}
	I_1&=\int_{-\infty}^{\infty}\int_{\sigma_2|x_2|/\sigma_1}^{\infty}x_1f_{T_1}(x_1)f_{T_2}(x_2)\mathrm{d}x_1\mathrm{d}x_2\nonumber\\
	&\stackrel{(a)}{=}\frac{2\alpha_{\nu_1}\alpha_{\nu_2}\nu_1}{\nu_1-1}\int\limits_{0}^{\infty}\Big(1+\frac{x_2^2\sigma_2^2}{\nu_1\sigma_1^2}\Big)^{\frac{1-\nu_1}{2}}\Big(1+\frac{x_2^2}{\nu_2}\Big)^{-\frac{\nu_2+1}{2}}\mathrm{d}x_2\nonumber\\
	&\stackrel{(b)}{=}\frac{\alpha_{\nu_1}\alpha_{\nu_2} B(\frac{1}{2},\frac{\nu_1+\nu_2-1}{2})}{(\nu_1-1)\nu_1^{-3/2}\sigma_2/\sigma_1}\ _2F_1\Big(\frac{\nu_2+1}{2},\frac{1}{2},\frac{\nu_1+\nu_2}{2},1-\frac{\nu_1\sigma_1^2}{\nu_2\sigma_2^2}\Big),\label{I1}
\end{align}
where (a) comes from solving the inner integral via \cite[eq. (2.27.7)]{Gradshteyn.2014}, while we leverage \cite[eq. (3.259.3)]{Gradshteyn.2014} to solve the remaining integral in (b).

In the case of $I_2$, we have that
\begin{align}
	I_2&\stackrel{(a)}{=}\int_{-\infty}^{\infty}x_2\big(1-F_{T_{1}}(-\sigma_2x_2/\sigma_1)\big)f_{T_{2}}(x_2)\mathrm{d}x_2\nonumber\\
	&\stackrel{(b)}{=}-\int_{-\infty}^0 x_2F_{T_{1}}(-\sigma_2x_2/\sigma_1)f_{T_{2}}(x_2)\mathrm{d}x_2 -\int_{0}^{\infty} x_2F_{T_{1}}(-\sigma_2x_2/\sigma_1)f_{T_{2}}(x_2)\mathrm{d}x_2\nonumber\\
	&\stackrel{(c)}{=}\int_{0}^{\infty} x_2F_{T_{1}}(\sigma_2x_2/\sigma_1)f_{T_{2}}(x_2)\mathrm{d}x_2 -\int_{0}^{\infty} x_2\big(1-F_{T_{1}}(\sigma_2x_2/\sigma_1)\big)f_{T_{2}}(x_2)\mathrm{d}x_2\nonumber\\
	&\stackrel{(d)}{=}2\underbrace{\int\limits_{0}^{\infty} x_2F_{T_{1}}\Big(\frac{\sigma_2x_2}{\sigma_1}\Big)f_{T_{2}}(x_2)\mathrm{d}x_2}_{I_{2,2}}-\underbrace{\int\limits_{0}^{\infty} x_2f_{T_{2}}(x_2)\mathrm{d}x_2}_{I_{2,1}},\label{I2}
\end{align}
where (a) comes from using the CDF definition, and (b) from leveraging $\int_{-\infty}^{\infty}x_2f_{T_2}(x_2)\mathrm{d}x_2=0$, followed by splitting the integration region such that the sign of $x_2$ can be fixed accordingly. The latter, together with the symmetry of $T_1$, is exploited to attain (c), while (d) is immediately obtained after simple algebraic transformations. Note that $I_{2}=2I_{2,2}-I_{2,1}$, where $I_1$ and $I_2$ require integral computations as shown in \eqref{I2}. Fortunately, their calculation can be further simplified as described next.

In the case of $I_{2,1}$, we have that
\begin{align}
	I_{2,1}&=\frac{\sqrt{\nu_2}\Gamma((\nu_2-1)/2)}{2\Gamma(\nu_2/2)\sqrt{\pi}},\label{I21}
\end{align}
which comes from using \cite[eq. (3.241.4)]{Gradshteyn.2014}. Meanwhile, 
\begin{align}
	I_{2,2}&=\int_{0}^{\infty} x_2F_{T_{1}}(\sigma_2x_2/\sigma_1)f_{T_{2}}(x_2)\mathrm{d}x_2\nonumber\\
	&\stackrel{(a)}{=} \int_{0}^{\infty} x_2\bigg(1-\frac{1}{2}\mathcal{I}_{\frac{\nu_1}{\frac{\sigma_2^2x_2^2}{\sigma_1^2}+\nu_1}}\Big(\frac{\nu_1}{2},\frac{1}{2}\Big)\bigg)f_{T_{2}}(x_2)\mathrm{d}x_2\nonumber\\
	&\stackrel{(b)}{=}\int\limits_{0}^{\infty} x_2f_{T_{2}}(x_2)\mathrm{d}x_2-\frac{\alpha_{\nu_2}}{2}\underbrace{\int_{0}^{\infty} x_2\mathcal{I}_{\frac{\nu_1}{\frac{\sigma_2^2x_2^2}{\sigma_1^2}+\nu_1}}\Big(\frac{\nu_1}{2},\frac{1}{2}\Big)\Big(1+\frac{x_2^2}{\nu_2}\Big)^{-\frac{\nu_2+1}{2}}\mathrm{d}x_2}_{I_{2,2}'}\nonumber\\
	&\stackrel{(c)}{=}I_{2,1}-\frac{\alpha_{\nu_2}}{2}I_{2,2}',\label{integral}
\end{align}
where (a) comes from substituting \eqref{FT}, (b) follows from rearranging terms, while (c) is obtained by leveraging \eqref{I21}. 

\begin{figure}[t!]
	\centering
	\includegraphics[width=0.75\textwidth]{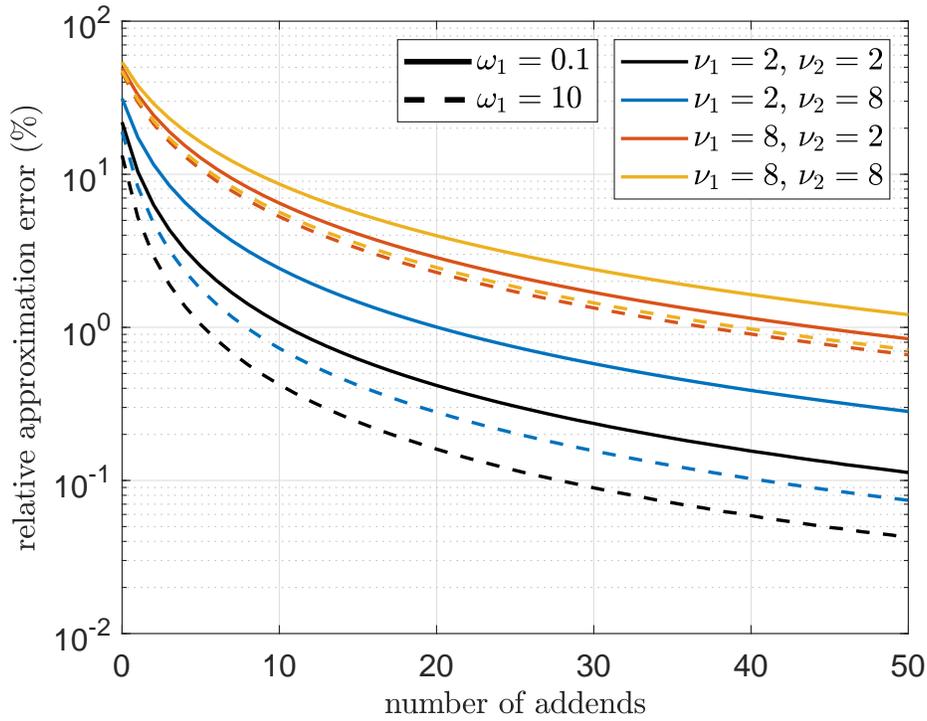}
	\caption{Accuracy of the computation of $I_{2,2}'$ according to \eqref{I22p2} with a finite number of addends.} 
	\label{Fig1}
\end{figure}

Now, to simplify $I_{2,2}'$, we introduce the following variable transformation
\begin{align}
	z^2\triangleq \frac{\nu_1}{\sigma_2^2x_2^2/\sigma_1^2+\nu_1}\ 
	\rightarrow\ \ x_2^2=\frac{\nu_1\sigma_1^2}{\sigma_2^2}\Big(\frac{1}{z^2}-1\Big),
\end{align}
thus, $x_2\mathrm{d}x_2=-\frac{\nu_1\sigma_1^2}{\sigma_2^2z^3}\mathrm{d}z$. 
Then, 
\begin{align}
	I_{2,2}'&=\omega_1\int_{0}^{1}z^{-3}\mathcal{I}_{z^2}\Big(\frac{\nu_1}{2},\frac{1}{2}\Big)\Big(\omega_2+\frac{1}{z^2}\Big)^{-\frac{\nu_2+1}{2}}\mathrm{d}z,\label{I22p}
\end{align}
where $\omega_1\triangleq \Big(\frac{\nu_1\sigma_1^2}{\sigma_2^2}\Big)^{-\frac{\nu_2-1}{2}}\nu_2^{\frac{\nu_2+1}{2}}$, $\omega_2\triangleq \frac{\nu_2\sigma_2^2}{\nu_1\sigma_1^2}-1$.  Now, we leverage the Taylor series expansion of an incomplete regularized beta function \cite{functions.wolfram.com} to write 
\begin{align}
	I_{z^2}\Big(\frac{\nu_1}{2},\frac{1}{2}\Big)&=\frac{2}{\sqrt{\pi}B\big(\frac{\nu_1}{2},\frac{1}{2}\big)}\sum_{i=0}^{\infty}\frac{ z^{\nu_1+2i}\Gamma(1/2+i)}{(\nu_1+2i)i!}.\label{Iz}
\end{align}
Then, by substituting \eqref{Iz}  into \eqref{I22p}, one gets
\begin{align}
	I_{2,2}'&=\frac{2\omega_1/\sqrt{\pi}}{B\big(\frac{\nu_1}{2},\frac{1}{2}\big)}\sum_{i=0}^{\infty}\frac{\Gamma\big(\frac{1}{2}+i\big)}{(\nu_1+2i)i!}\int\limits_{0}^{1}\frac{\big(\omega_2+\frac{1}{z^2}\big)^{-\frac{\nu_2+1}{2}}}{z^{3-\nu_1-2i}}\mathrm{d}z\nonumber\\
	&\stackrel{(a)}{=}\frac{\omega_1/\sqrt{\pi}}{B\big(\frac{\nu_1}{2},\frac{1}{2}\big)}\sum_{i=0}^{\infty}\frac{\Gamma\big(\frac{1}{2}+i\big)}{(\nu_1+2i)i!}\int\limits_{0}^{1}\frac{\big(y\omega_2+1\big)^{-\frac{\nu_2+1}{2}}}{y^{\frac{3-\nu_1-\nu_2-2i}{2}}}\mathrm{d}y\nonumber\\
	&\stackrel{(b)}{=}\frac{2\omega_1}{\sqrt{\pi}B\big(\frac{\nu_1}{2},\frac{1}{2}\big)}\sum_{i=0}^{\infty}\frac{\Gamma\big(i+\frac{1}{2}\big)\ _2F_1\big(\frac{\nu_2+1}{2},\frac{\nu_1+\nu_2+2i-1}{2},\frac{\nu_1+\nu_2+2i+1}{2},-\omega_2\big)}{(\nu_1+2i)(\nu_1+\nu_2+2i-1)i!},\label{I22p2}
\end{align}
where (a) comes from exploiting the variable transformation $y=z^2$,  while the integral is solved in  (b) by leveraging \cite[eq. (3.197.8)]{Gradshteyn.2014}. Then $I_{2,2}'$ can be easily estimated by truncating the infinite sum in \eqref{I22p2}. As illustrated in Figure~\ref{Fig1}, the relative approximation error decreases following a power law decay, thus, a relatively small number of addends is needed, especially for small $\nu_2$, $\nu_1$, and large $\omega_2$.

By combining \eqref{I1}, \eqref{I2}, \eqref{I21}, \eqref{integral}, \eqref{I22p2}, and substituting them into \eqref{EZ}, we obtain 
\begin{align}
	\mathbb{E}[|Z|]=&
	\frac{ B(\frac{1}{2},\frac{\nu_1+\nu_2-1}{2})\ _2F_1\Big(\frac{\nu_2+1}{2},\frac{1}{2},\frac{\nu_1+\nu_2}{2},1-\frac{\nu_1\sigma_1^2}{\nu_2\sigma_2^2}\Big)}{(\nu_1-1)\nu_1^{-3/2}\sigma_2/\big(2 \alpha_{\nu_1}\alpha_{\nu_2}\sigma_1^2\big)} + \frac{\sigma_2\Gamma(\frac{\nu_2-1}{2})}{\Gamma(\frac{\nu_2}{2})\sqrt{\frac{\pi}{\nu_2}}}- \frac{4\alpha_{\nu_2}\sigma_2\omega_1}{\sqrt{\pi}B\big(\frac{\nu_1}{2},\frac{1}{2}\big)}\nonumber\\
	&\qquad\qquad \times\sum_{i=0}^{\infty}\frac{\Gamma\big(i+\frac{1}{2}\big)\ _2F_1\big(\frac{\nu_2+1}{2},\frac{\nu_1+\nu_2+2i-1}{2},\frac{\nu_1+\nu_2+2i+1}{2},-\omega_2\big)}{(\nu_1+2i)(\nu_1+\nu_2+2i-1)i!}. \label{EZ2}
\end{align}
\subsection{Special case: Sum of i.i.d. Student's $t$ RVs}
Herein, we consider the special case of the sum of i.i.d. Student's $t$ RVs, i.e., $\sigma_i=\sigma$, $\nu_i=\nu$, $\forall i=1,\cdots,K$. With this in hand,  the computation of $\mathbb{E}[Z^2]$, $\mathrm{CF}_Z(r)$, and $\mathbb{E}[|Z|]$ (for $K=2$) can be more easily obtained as follows.

Under such conditions, \eqref{Z1} and \eqref{CF2} directly simplify  to
\begin{align}
	\mathbb{E}[Z^2]&=\frac{\sigma\nu K}{\nu-2},\\
	\mathrm{CF}_Z(r)&=\mathrm{CF}_{T}^K(\sigma r)=\bigg[\frac{(\sqrt{\nu}\sigma|r|)^{\nu/2}K_{\nu/2}(\sqrt{\nu}\sigma|r|)}{2^{\nu/2-1}\Gamma(\nu/2)}\bigg]^K.
\end{align}
As for $\mathbb{E}[|Z|]$ for $K=2$, for which $\mathbb{E}[|Z|]=2\sigma(I_1+I_{2,1}-\alpha_{\nu} I_{2,2}')$, one may depart from (a) in \eqref{I1} to write 
\begin{align}
	I_1= \frac{2\alpha_\nu^2 \nu}{\nu-1}\int\limits_0^\infty\Big(1+\frac{x_2^2}{\nu}\Big)^{-\nu}\mathrm{d}x_2=\frac{\Gamma\Big(\frac{\nu-1}{2}\Big)\Gamma\Big(\nu-\frac{1}{2}\Big)}{2^\nu \Gamma(\nu/2)^{3}\nu^{-1/2}},
\end{align}
which leverages \cite[eq. (3.251.2)]{Gradshteyn.2014} and \eqref{anu} followed by some simple algebraic simplifications.
Meanwhile, \eqref{I21} can be directly used with $\nu$ instead of $\nu_2$. Now, \eqref{I22p2} can be computed by departing from \eqref{I22p} as
\begin{align}
	I_{2,2}'&= \nu \int_0^1 z^{\nu-2}\mathcal{I}_{z^2}\Big(\frac{\nu}{2},\frac{1}{2}\Big)\mathrm{d}z\nonumber\\
	&\stackrel{(a)}{=} \frac{\Gamma\big(\frac{\nu+1}{2}\big)}{\Gamma(\nu/2)\sqrt{\pi}} \int_0^1 x^{\nu-3/2} \ _2F_1\Big(\frac{\nu}{2},\frac{1}{2},1+\frac{\nu}{2},x\Big)\mathrm{d}x\nonumber\\
	&\stackrel{(b)}{=} \frac{\nu}{\nu-1}\bigg(1-\frac{\sqrt{\pi}\Gamma\big(\nu-\frac{1}{2}\big)}{2^{\nu-1}\Gamma(\nu/2)^2}\bigg),
\end{align}
where (a) comes from stating the regularized incomplete beta function in terms of a hypergeometric function according to \cite[eq. (8.17.7)]{Olver.2010} and using $x\triangleq z^2\rightarrow \mathrm{d}x=2z\mathrm{d}z$, while (b) follows from leveraging \cite{functions.wolfram2.com} together with iterative integration by parts. Finally, by combining the above results, we obtain
\begin{align}
	\mathbb{E}[|Z|]=\frac{\sigma\sqrt{\nu}\Gamma\big(\frac{\nu-1}{2}\big)\Gamma\big(\nu-\frac{1}{2}\big)}{2^{\nu-2}\Gamma(\nu/2)^3}.
\end{align}

\section{Distribution Fitting}\label{main}
Next, we follow two different distribution fitting approaches. The first approach is based on matching the second and absolute moments, while the second approach relies on matching the second moment and the CF for a certain $r$. We also illustrate their accuracy.
\subsection{Fitting based on Second and Absolute Moments}\label{FSA}
\begin{figure}[t!]
	\centering
	\includegraphics[width=0.75\textwidth]{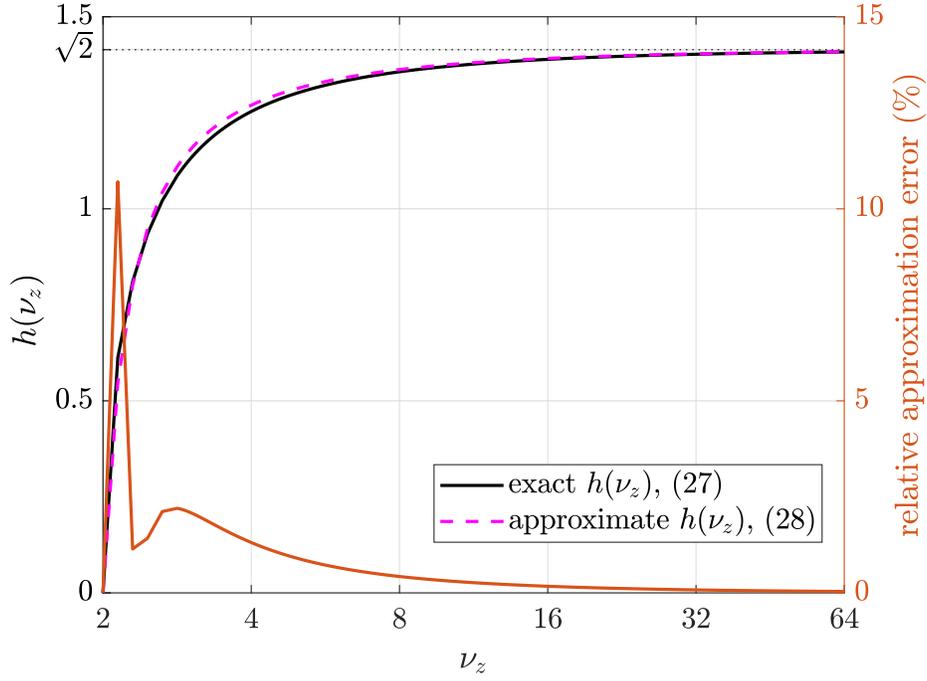}
	\caption{Accuracy of the approximate computation of $h(\nu_z)$ given in \eqref{happ}.}
	\label{Fig2}
\end{figure}

According to \eqref{Er} with $m=2$ and \eqref{abs} with $m=1$, the set of equations to solve is 
\begin{align}
	\left\{\frac{\sigma_z^2\nu_z}{\nu_z-2}=\mathbb{E}[Z^2],\qquad 
	\frac{\sigma_z\sqrt{\nu_z}\Gamma((\nu_z-1)/2)}{\Gamma(\nu_z/2)\sqrt{\pi}}=\mathbb{E}[|Z|]	\right\}\label{system}
\end{align}
with variables $\{\sigma_z,\nu_z\}$. Recall that $\mathbb{E}[Z^2]$ is given by \eqref{Z1}, while $\mathbb{E}[|Z|]$ is given in \eqref{EZ2} for the case of $K=2$. By isolating $\sigma_z$ in the first equation, i.e., $\sigma_z=\sqrt{(\nu_z-2)\mathbb{E}[Z^2]/\nu_z}$, and substituting it into the second equation, the system of equations \eqref{system} transforms to
\begin{align}
	h(\nu_z)\triangleq\frac{\Gamma((\nu_z-1)/2)\sqrt{\nu_z-2}}{\Gamma(\nu_z/2)}&=\frac{\sqrt{\pi}\mathbb{E}[|Z|]}{\sqrt{\mathbb{E}[Z^2]}}.\label{equation}
\end{align}
Observe that attaining an exact closed-form solution for $\nu_z$ in \eqref{equation} is not viable, thus, we resort to a low-complex approximation of $h(\nu_z)$. For this, we plot $h(\nu_z)$ vs $\nu_z$ in Figure~\ref{Fig2}, and realize that $h(\nu_z)$ has approximately the form of a quotient of two linear functions. Hence, we state
\begin{align}
	h(\nu_z)\approx \frac{p_1\nu_z + p_2}{\nu_z+p_3}.\label{happ}
\end{align}
Here, we know that $h(2)=0$, and
\begin{align}
	\lim_{\nu_z\rightarrow\infty}h(\nu_z)&=\lim_{\nu_z\rightarrow\infty}\frac{\Gamma((\nu_z-1)/2)\sqrt{\nu_z-2}}{\Gamma(\nu_z/2)} \stackrel{(a)}{=}\lim_{\nu_z\rightarrow\infty}\sqrt{\frac{2(\nu_z-2)}{\nu_z}}=\sqrt{2},
\end{align}
where (a) comes from using \cite[eq. 5.11.12]{Olver.2010}. Then, in order to satisfy such conditions,
we can directly set $p_1$ and $p_2$ as follows
\begin{align}
	\lim_{\nu_z\rightarrow\infty}h(\nu_z)&=p_1=\sqrt{2},\label{a1}\\
	h(2)&=\frac{2p_1+p_2}{2+p_3}=0 \rightarrow 2p_1+p_2=0\rightarrow p_2 =-2\sqrt{2},\label{a2}
\end{align}
where \eqref{a2} uses the result in \eqref{a1} in the last step. 
Finally, $p_3= -\sqrt{3}$ can be obtained easily by standard curve fitting. The accuracy of \eqref{happ} is also depicted in Figure~\ref{Fig2}.

\begin{figure}[t!]
	\centering
	\includegraphics[width=0.75\textwidth]{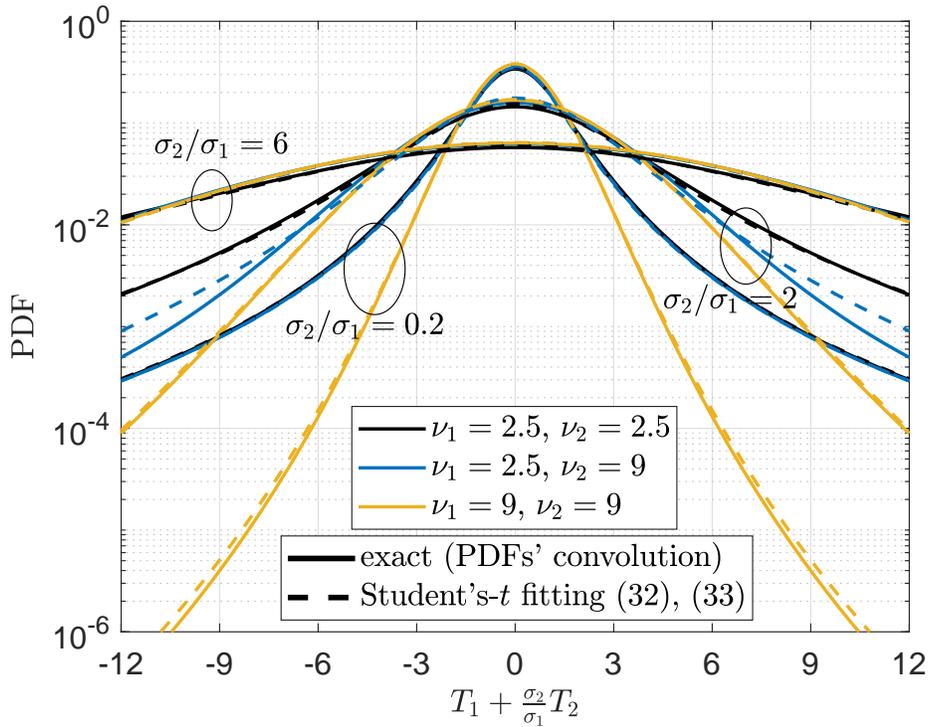}
	\caption{PDF of the linear combination of two Student's $t$ RVs.} 
	\label{Fig3}
\end{figure}

With \eqref{happ} in place, one can estimate $\nu_z$ and $\sigma_z$ as
\begin{align}
	\nu_z^\star&\approx \frac{\pi\mathbb{E}[|Z|]-2\sqrt{2\mathbb{E}[Z^2]}}{\sqrt{\pi}\mathbb{E}[|Z|]-\sqrt{2\mathbb{E}[Z^2]}},\label{nuz}\\
	\sigma_z^\star&\approx \sqrt{\frac{(\pi-2\sqrt{\pi})\mathbb{E}[|Z|]\mathbb{E}[Z^2]}{\pi\mathbb{E}[|Z|]-2\sqrt{2\mathbb{E}[Z^2]}}},\label{sigmaz}
\end{align}
and use the approximation $Z\sim \sigma_z^\star\mathcal{T}(\nu_z^\star)$. Such a distribution fitting is illustrated in Figure~\ref{Fig3} and evinces the appropriateness of our  approach. All in all, the fitting accuracy only seems to be affected in the distribution tails, which is expected considering that only two (low-order) features of the child distributions are used for the fitting. Nevertheless, even in the tails region, the accuracy is surprisingly good, being only critically affected when the child distributions have significantly diverging degrees of freedom and scaling factors in opposite directions, e.g., small $\nu_1,\sigma_2$ and large $\nu_2,\sigma_1$.

\subsubsection{Linear Combination of $K>2$ Student's $t$ RVs}
Notice that if a scaled Student's $t$ distribution fits accurately the distribution of the linear combination of $K=2$ Student's $t$ RVs, then such a Student's $t$ fitting approach applies for the linear combination of any $K\ge 2$ Student's $t$ RVs. This can be easily shown by induction as follows. 

According to the previous subsection's results, we can state that
\begin{align}
	\sigma_1T_1+\sigma_2T_2\sim \sigma_{z_2} \mathcal{T}(\nu_{z_2})
\end{align}
holds approximately. Here, $\sigma_{z_n}=g_1(\sigma_1,\cdots,\sigma_n,\nu_1,\cdots,\nu_n)$ and $\nu_{z_n}=g_2(\sigma_1,\cdots,\sigma_n,\nu_1,\cdots,\nu_n)$, where $g_1(\cdot)$ and $g_2(\cdot)$ are transformation functions (given by \eqref{nuz} and \eqref{sigmaz} in the particular case of $n=2$). Then, assume that $\sum_{i=1}^K\sigma_iT_i\sim \sigma_{z_{K}}\mathcal{T}(\nu_{z_K})$, and observe that
\begin{align}
	\sum_{i=1}^{K+1}\sigma_iT_i&= \sigma_{K+1}\underbrace{T_{K+1}}_{\mathcal{T}(\nu_{K+1})}+\underbrace{\sum_{i=1}^{K}\sigma_iT_i}_{\sigma_{z_{K}}\mathcal{T}(\nu_{z_K})}\nonumber\\
	&\sim \underbrace{g_{1}(\sigma_{z_{K}},\sigma_{K+1},\nu_{z_K},\nu_{K+1})}_{\sigma_{z_{K+1}}} \mathcal{T}\big(\underbrace{g_{1}(\sigma_{z_{K}},\sigma_{K+1},\nu_{z_K},\nu_{K+1})}_{\nu_{z_{K+1}}}\big).
\end{align}
Thus, proving the hypothesis.

\begin{figure}[t!]
	\centering
	\includegraphics[width=0.75\textwidth]{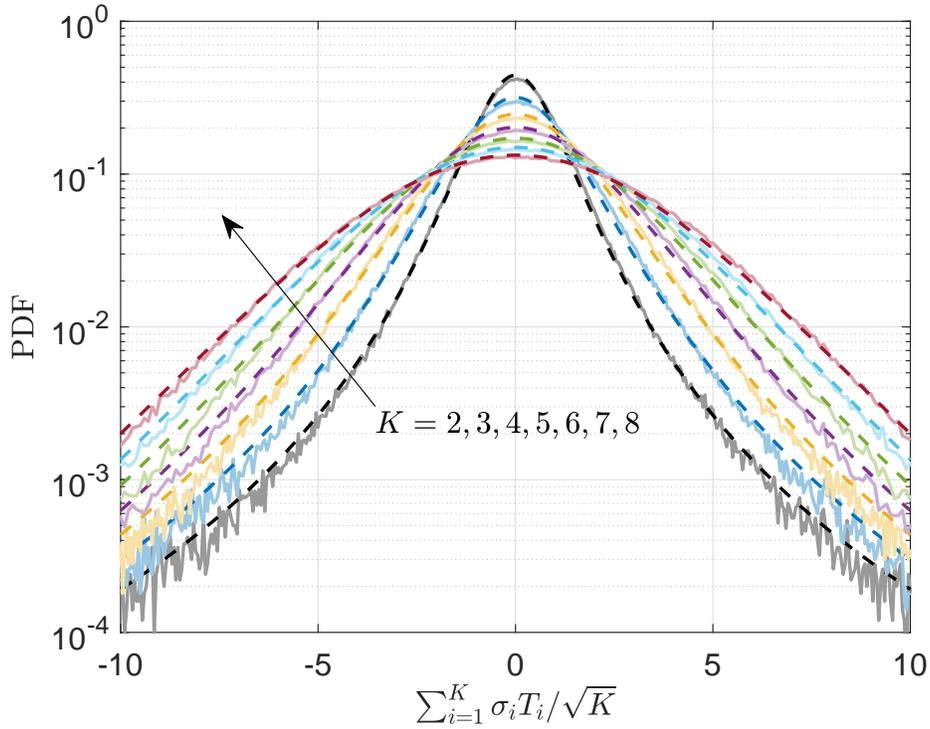}
	\caption{PDF of the (normalized) linear combination of $K$ Student's $t$ RVs. We set $\sigma_i=i/2$ and $\nu_i=2+i/2.$ Straight lines correspond to the empirical PDF obtained through Monte Carlo simulations, while dotted lines correspond to our proposed Student's $t$ distribution fitting based on second and absolute moments matching.} 
	\label{Fig4}
\end{figure}

Wrapping up, the linear combination of $K\ge 2$ Student's $t$ RVs is approximately distributed as $\sigma_{z_K}\mathcal{T}(\nu_{z_K})$, where $\sigma_{z_K}$, $\nu_{z_K}$ can be iteratively obtained as
\begin{align}
	\sigma_{z_{i+1}}=g_1(\sigma_{z_i},\sigma_{i+1},\nu_{z_i},\nu_{i+1}),\\
	\nu_{z_{i+1}}=g_2(\sigma_{z_i},\sigma_{i+1},\nu_{z_i},\nu_{i+1}),
\end{align}
$\forall i\ge 1$, where $\sigma_{z_1}=\sigma_1$, $\nu_{z_1}=\nu_1$, and $g_1(\sigma_1,\sigma_2,\nu_1,\nu_2)$, $g_2(\sigma_1,\sigma_2,\nu_1,\nu_2)$ are respectively given by \eqref{nuz} and \eqref{sigmaz}, which can be computed by leveraging \eqref{Z1}, \eqref{EZ2}. The accuracy of such a procedure has been corroborated by several simulation campaigns, and it is illustrated here in Figure~\ref{Fig4} for an example set of distribution parameters.

\subsection{Fitting based on Second Moment and Characteristic Function}\label{M2C}
\begin{figure}[t!]
	\centering
	\includegraphics[width=0.75\textwidth]{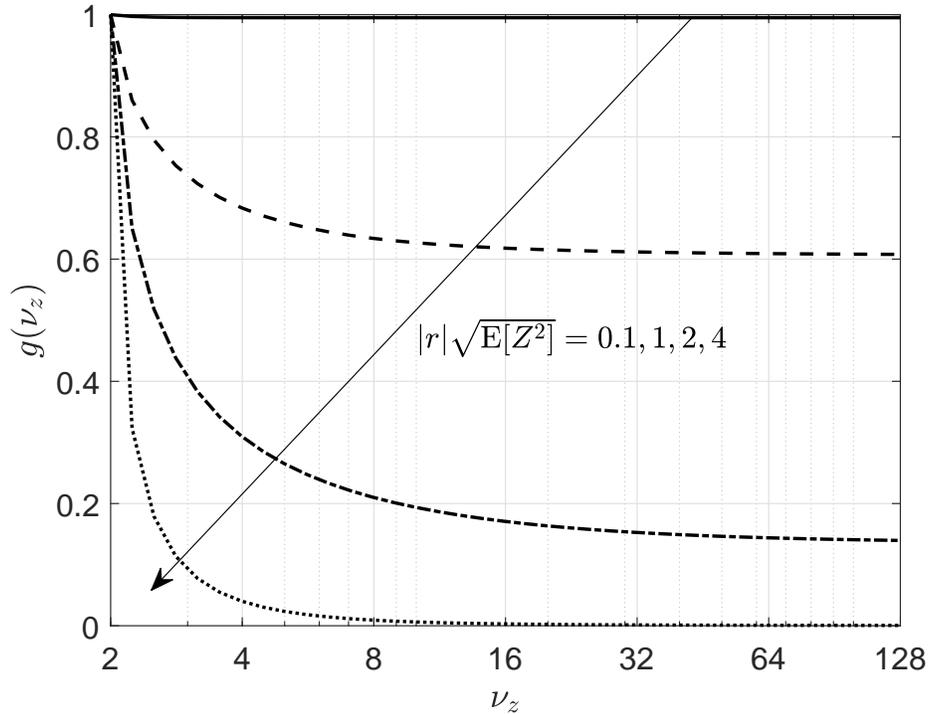}
	\caption{$g(\nu_z)$ vs $\nu_z$ for $|r|\sqrt{\mathbb{E}[Z^2]}\in\{0.1,1,2,4\}$.} 
	\label{Fig5}
\end{figure}
According to \eqref{Er} with $m=2$ and \eqref{CF}, the set of equations to solve is 
\begin{align}
	\left\{\frac{\sigma_z^2\nu_z}{\nu_z-2}=\mathbb{E}[Z^2],\qquad 
	\frac{(\sqrt{\nu_z}\sigma_z|r|)^{\nu_z/2}K_{\nu_z/2}(\sqrt{\nu_z}\sigma_z|r|)}{2^{\nu_z/2-1}\Gamma(\nu_z/2)}=\mathrm{CF}_Z(r)	\right\}\label{system2}
\end{align}
with variables $\{\sigma_z,\nu_z\}$. Recall that $\mathbb{E}[Z^2]$ is given by \eqref{Z1}, while $\mathrm{CF}_Z(r)$ is given in \eqref{CF2}. As in Section~\ref{FSA}, we first isolate $\sigma_z$ in the first equation, i.e., $\sigma_z=\sqrt{(\nu_z-2)\mathbb{E}[Z^2]/\nu_z}$, and then substitute it into the second equation. By doing this, the system of equations \eqref{system2} transforms to
\begin{align}
	g(\nu_z)\triangleq\frac{\big(|r|\sqrt{(\nu_z-2)\mathbb{E}[Z^2]}\big)^{\nu_z/2}K_{\nu_z/2}(|r|\sqrt{(\nu_z-2)\mathbb{E}[Z^2]})}{2^{\nu_z/2-1}\Gamma(\nu_z/2)}&=\mathrm{CF}_Z(r).\label{equation2}
\end{align}
It can be shown that $g(\nu_z)$ is a decreasing function and that $\lim_{\nu_z\rightarrow 2^+}g(\nu_z)=1$ and  $\lim_{\nu_z\rightarrow \infty}g(\nu_z)=l(|r|\sqrt{\mathbb{E}[Z^2]})\ge 0$ for some decreasing function $l(\cdot)$. This is illustrated in Figure~\ref{Fig5} and implies that there is at most one real solution $\nu_z^\star$ for \eqref{equation2}, which can be easily found via the bisection method. After this, one sets $\sigma_z^\star=\sqrt{(\nu_z^\star-2)\mathbb{E}[Z^2]/\nu_z^\star}$ and uses the approximation $Z\sim \sigma_z^\star\mathcal{T}(\nu_z^\star)$.

The main challenge with this approach is the proper setting of $r$. On one hand, \eqref{equation2} may not have solution for certain values of $r$. On the other hand, different feasible values of $r$ may lead to significantly different fitting accuracy figures. We investigate these issues in Section~\ref{paramCF}. For now, let us conveniently set %
\begin{align}
	r=\mathbb{E}[Z^2]^{-1/2} \label{rset}
\end{align}
so that $g(\nu_z)$ simplifies to
\begin{align}
	g(\nu_z)\triangleq\frac{(\nu_z-2)^{\nu_z/4}K_{\nu_z/2}(\sqrt{\nu_z-2})}{2^{\nu_z/2-1}\Gamma(\nu_z/2)}. \label{gap}
\end{align}
Observe that attaining an exact closed-form solution for $\nu_z$ in $g(\nu_z)=\mathrm{CF}_Z(\mathbb{E}[Z^2]^{-1/2})$ using \eqref{gap} is still not viable. Fortunately, $g(\nu_z)$ can be accurately approximated by a very tractable fractional function of the form
\begin{align}
	g(\nu_z)\approx \frac{p_1\nu_z + p_2}{\nu_z+p_3}.\label{gapp}
\end{align}
Meanwhile, we know that $\lim_{\nu_z\rightarrow 2^+}g(\nu_z)=1$, and $\lim_{\nu_z\rightarrow \infty}g(\nu_z)\approx 0.6070$ as shown in Figure~\ref{Fig5}. Then, in order to satisfy such conditions, we can directly set $p_1$ and $p_2$ as follows
\begin{figure}[t!]
	\centering
	\includegraphics[width=0.75\textwidth]{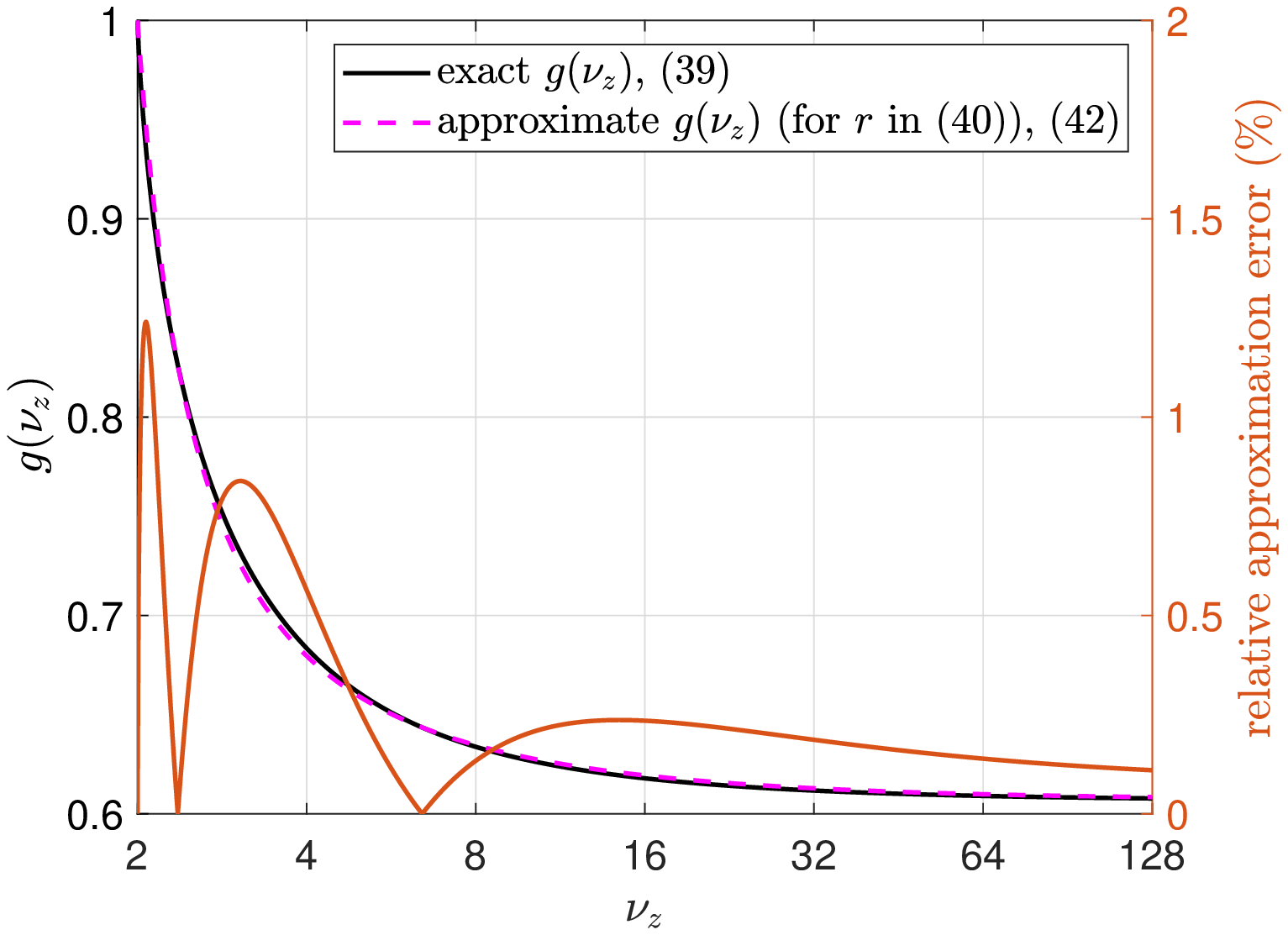}
	\caption{Accuracy of the approximate computation of $g(\nu_z)$ given in \eqref{gapp}.} 
	\label{Fig6}
\end{figure}
\begin{figure}[t!]
	\centering
	\includegraphics[width=0.75\textwidth]{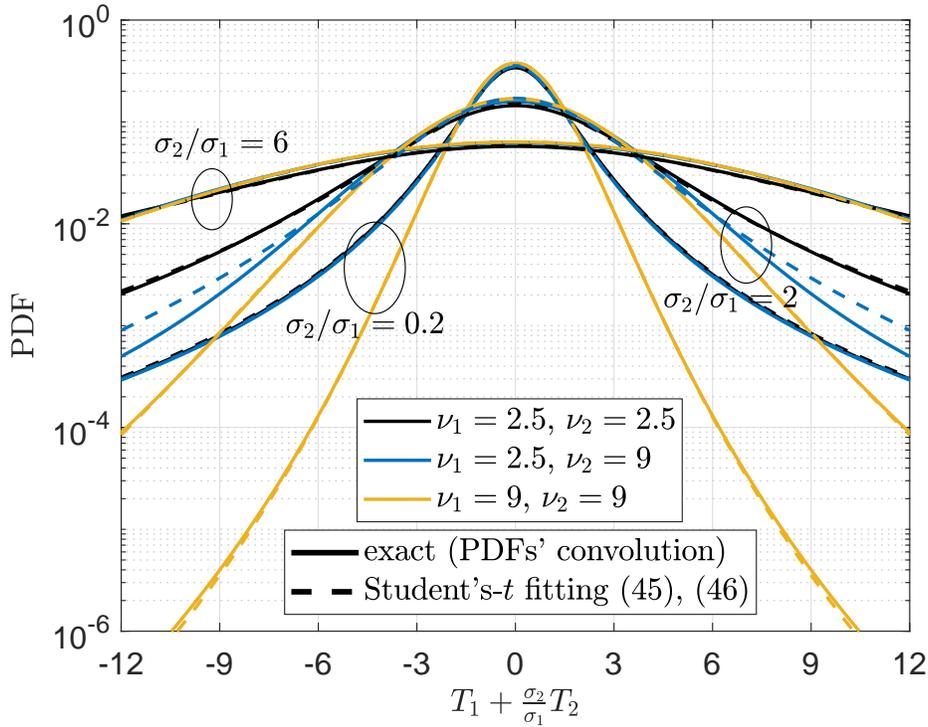}
	\caption{PDF of the linear combination of two Student's $t$ RVs.} 
	\label{Fig7}
\end{figure}
\begin{align}
	\lim_{\nu_z\rightarrow\infty}g(\nu_z)&=p_1=0.607,\label{a12}\\
	g(2)&\approx \frac{2p_1+p_2}{2+p_3}=1 \rightarrow 2p_1+p_2=2+p_3\rightarrow p_3 =2p_1+p_2-2=p_2-0.786,\label{a22}
\end{align}
where \eqref{a22} leverages the result in \eqref{a12} in the last step. Finally, $p_2=-0.7606$ can be obtained easily by standard curve fitting, and then one can set $p_3=-0.7606-0.786=-1.5466$. The accuracy of \eqref{gapp} is depicted in Figure~\ref{Fig6}.

\begin{figure}[t!]
	\centering
	\includegraphics[width=0.75\textwidth]{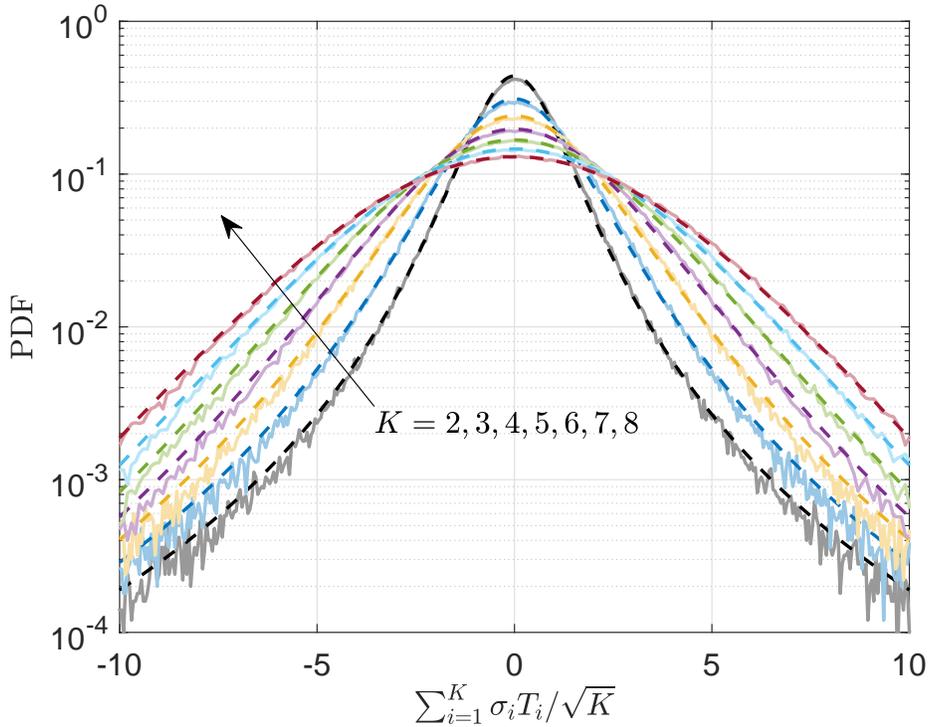}
	\caption{PDF of the (normalized) linear combination of $K$ Student's $t$ RVs. We set $\sigma_i=i/2$ and $\nu_i=2+i/2.$ Straight lines correspond to the empirical PDF obtained through Monte Carlo simulations, while dotted lines correspond to our proposed Student's $t$ distribution fitting based on the matching of the second moments and the CFs for $r$ given in \eqref{rset}.} 
	\label{Fig8}
\end{figure}
With \eqref{gapp} in place, one can estimate $\nu_z$ and $\sigma_z$ as
\begin{align}
	\nu_z^\star&\approx \frac{0.7606-1.5466\times\mathrm{CF}_Z(\mathbb{E}[Z^2]^{-1/2})}{0.607-\mathrm{CF}_Z(\mathbb{E}[Z^2]^{-1/2})},\label{nuz2}\\
	\sigma_z^\star&\approx \sqrt{\frac{(\mathrm{CF}_Z(\mathbb{E}[Z^2]^{-1/2})-1)\mathbb{E}[Z^2]}{1.6775-3.4111\times \mathrm{CF}_Z(\mathbb{E}[Z^2]^{-1/2})}},\label{sigmaz2}
\end{align}
and use the approximation $Z\sim \sigma_z^\star\mathcal{T}(\nu_z^\star)$. Such a distribution fitting is illustrated in Figure~\ref{Fig7} and Figure~\ref{Fig8}, and evinces the appropriateness of our  approach. Finally, notice that the results here agree also with our previous observations around Figure~\ref{Fig3}.
\section{Fitting Accuracy Analysis}\label{fitting}
In this section, we assess the accuracy of the fitting methods discussed in Section~\ref{main} by adopting the Bhattacharyya distance metric \cite{Kailath.1967}. This metric measures the similarity of two probability distributions, which in this case are the true distribution of $Z$, $f_Z(z)$, and the approximate scaled Student's $t$ distribution $f_{\hat{Z}}(z)$ based on one of the proposed fitting approaches.  Notice that since the true/exact distribution of $Z$, $f_Z(z)$, is unknown and difficult to compute for a general $K$, we leverage a Monte Carlo approach to estimate the Bhattacharyya distance. Specifically, such a metric is computed as follows
\begin{align}
	d_B(Z,\hat{Z})=-\ln\Big(\sum_{n=1}^{N}\sqrt{f_Z(z_n)f_{\hat{Z}}(z_n)}\Big),\label{KL}
\end{align}
where $z_n$ is the $n-$th sample taken from $\sigma\sum_{i=1}^KT_i$, and $f_Z(z_n)$ is estimated using its histogram. As $N\rightarrow\infty$, \eqref{KL} approaches the exact Bhattacharyya distance between the continuous probability distributions of $Z$ and $\hat{Z}$.
Finally, we focus on the special case of the sum of i.i.d. RVs for simplicity, adopt $N=10^6$, and set $\sigma=1$ without loss of generality.
\subsection{Absolute Moment vs CF -based Fitting}
\begin{figure*}[t!]
	\centering
	\includegraphics[width=0.77\textwidth]{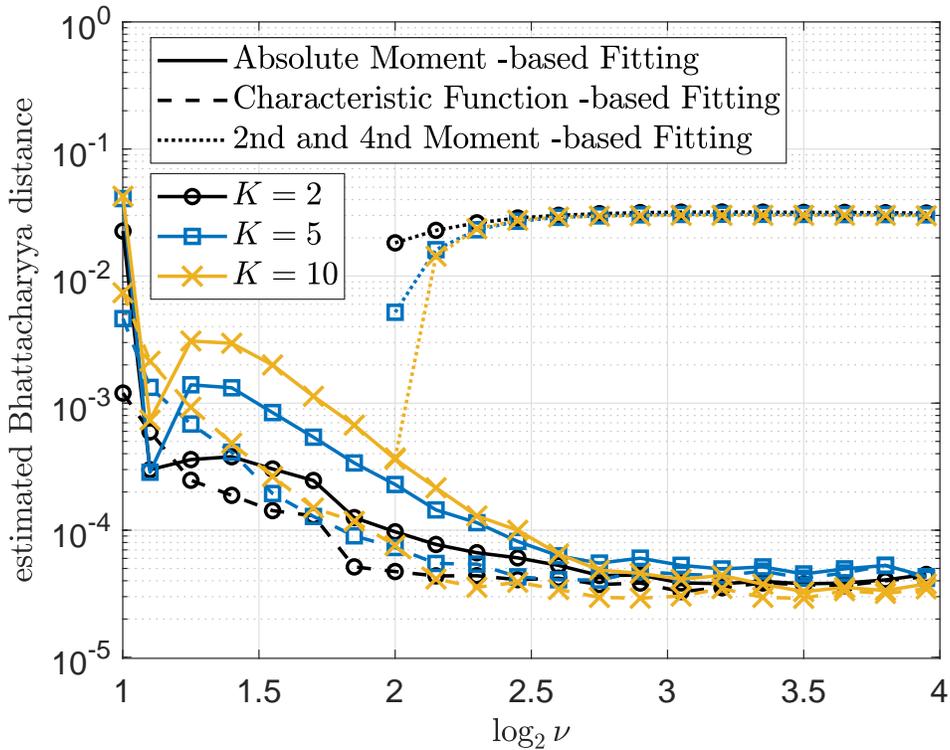}
	\caption{Estimated Bhattacharyya distance as a function of $\nu$.} 
	\label{Fig9}
\end{figure*}

Figure~\ref{Fig9} shows the fitting accuracy of the two methods proposed in this work together with that of a benchmark approach based on second and fourth moments matching. As commented earlier, such an approach can only be used when $\nu\ge 4$, which was one of the key motivations for our work. From the figure, we can tell that  
i) the proposed fitting approaches are more accurate than the benchmark based on second and fourth moments matching, which is also restricted to the cases where $\nu\ge 4$; 
ii) the fitting based on the CF matching is generally more accurate than the one based on absolute moment matching, although both approaches tend to converge as $\nu$ increases; and 
iii) for relatively small $\nu$, the proposed fitting approaches are more accurate for a smaller $K$, while this behavior might be reverted as $\nu$ increases.

\subsection{Scaling Laws of the Fitting Parameters}\label{scaling}
\begin{figure*}[t!]
	\centering
	\includegraphics[width=0.77\textwidth]{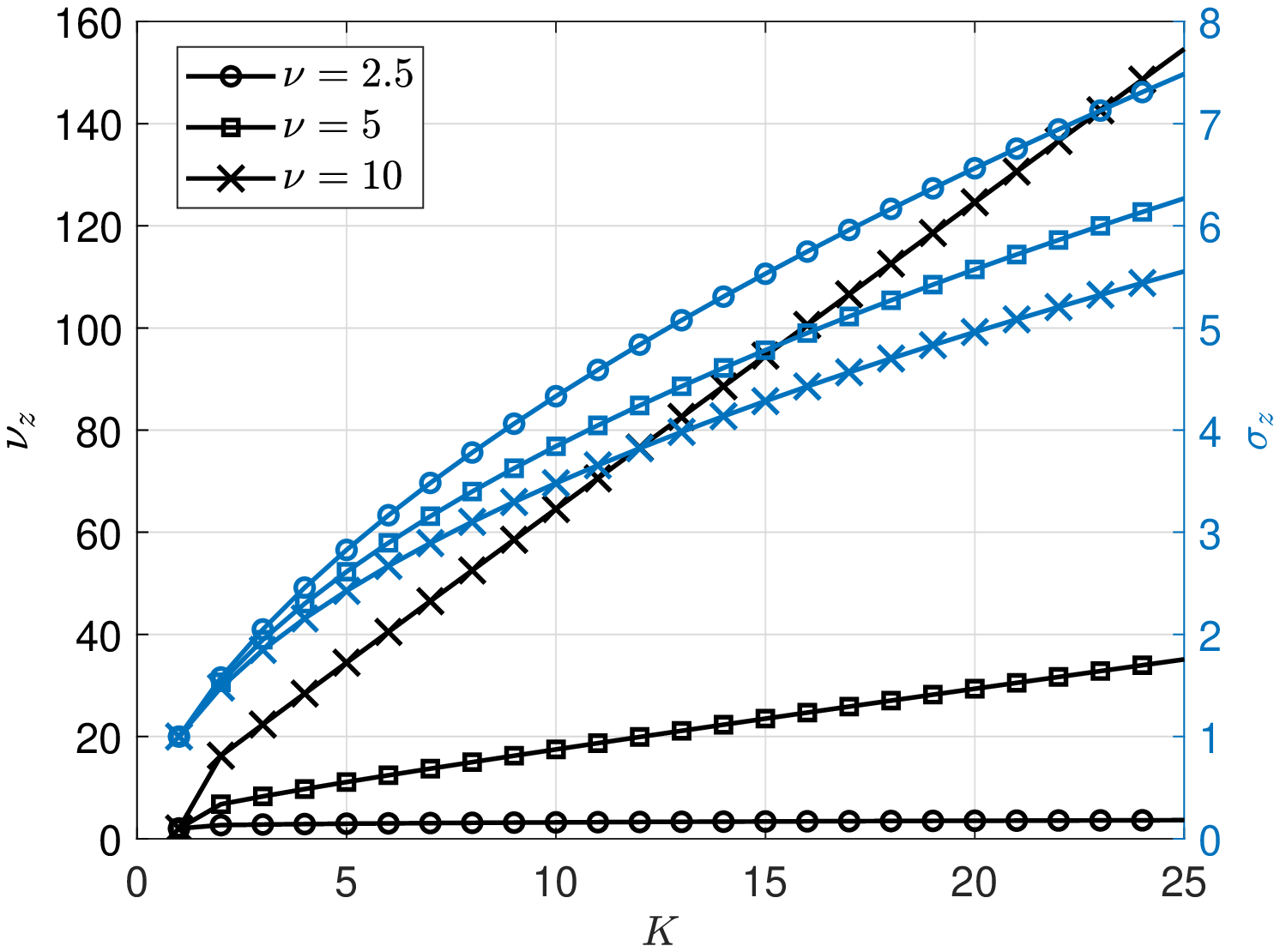}
	\caption{Parameters of the CF-based fitting distribution  as a function of $K$ for $\nu\in\{2.5,5,10\}$.} 
	\label{Fig10}
\end{figure*}

Herein, we leverage the i.i.d. assumption to illustrate in Figure~\ref{Fig10} how the parameters of the fitting distribution scale with $K$. For this, we adopt only the CF-based fitting as our previous results indicate that it is the most accurate. 
Observe that both parameters of the fitting distribution, the scale $\sigma_z$ and the number of degrees of freedom $\nu_z$, increase with $K$. Specifically, $\sigma_z,\nu_z \sim \gamma_1 K^{\gamma_2}+\gamma_3$ with $\gamma_1,\gamma_2 >0$ and $\gamma_2<1$ (concave increase). Moreover, $\sigma_z$ and $\nu_z$ are respectively increasing and decreasing functions of $\nu$.
\begin{figure*}[t!]
	\centering
	\includegraphics[width=0.77\textwidth]{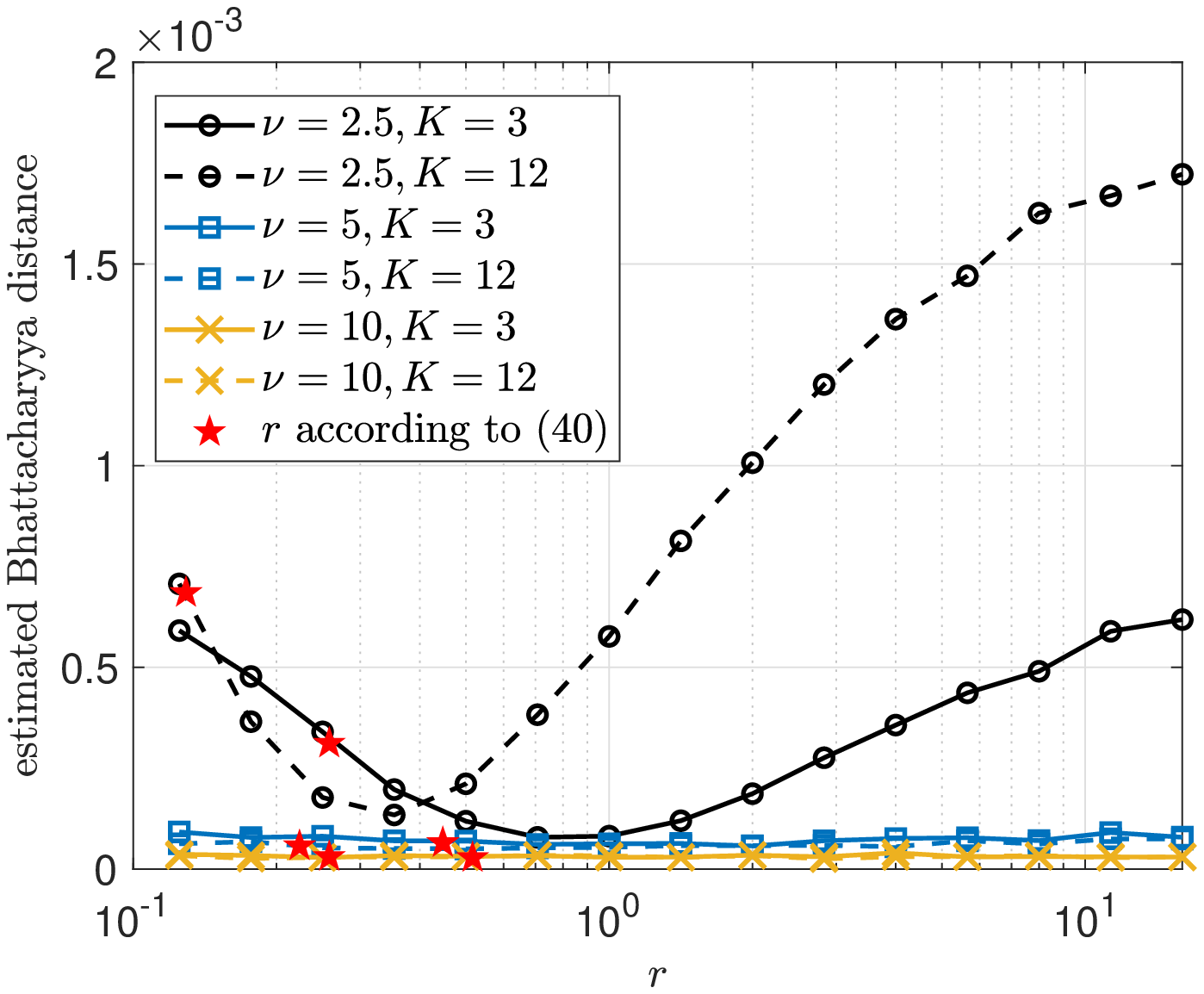}
	\caption{Estimated Bhattacharyya distance under the CF-based fitting as a function of $r$ for $\nu\in\{2.5,5,10\}$ and $K\in\{3,12\}$.} 
	\label{Fig11}
\end{figure*}
\subsection{Performance Impact of the CF Parameter Setting}\label{paramCF}
As commented earlier in Section~\ref{M2C}, different choices of $r$ in \eqref{system2} (or directly in \eqref{equation2}) may lead to significantly different fitting accuracy figures. Here, we investigate these issues by illustrating the estimated Bhattacharyya distance as a function of $r$ in Figure~\ref{Fig11}. Observe that for relatively large values of $\nu$, the accuracy depends little on the specific value of $r$. This situation changes drastically as $\nu$ decreases approaching two, under which properly setting $r$ becomes more critical. Indeed, there is an accuracy-optimum value of $r$, especially noticeable when $\nu$ is small. Notice that although the optimum $r$ may be cumbersome to determine beforehand in practice. Nevertheless, and as a rule of dumb, relatively small values of $r$ are usually preferred, and our proposal in \eqref{rset} appears to be a valid (and simple) configuration approach.

\section{Conclusion}\label{conclusions}
In this work, we proposed two fitting approaches for the distribution of linear combinations of Student's $t$ RVs with more than two degrees of freedom. They  leverage the second moment together with either the first absolute moment or the CF to fit the distribution to that of a scaled Student's $t$ RV. For the former, we first analytically obtained the absolute moment of a linear combination of $K= 2$ Student's $t$ RVs and then generalized to $K\ge 2$ through a simple iterative procedure, while the fitting is direct for the latter but its accuracy depends on the CF parameter. Notably, we proposed a simple CF parameter configuration and showed that it can lead to high fitting accuracy. We resorted to Monte Carlo simulations and adopted the Bhattacharyya distance metric for numerically quantifying the fitting accuracy. We showed that the CF-based fitting can usually outperform the absolute moment -based fitting, although the accuracy provided by both approaches converges when the $t-$RVs have a sufficiently large number of degrees of freedom. Interestingly, both proposed approaches outperform a benchmark fitting based on second and fourth moments matching, which is only applicable when all $t-$RVs have at least four degrees of freedom. Interestingly, both the scale and number of degrees of freedom of the fitting distribution where shown to increase almost linearly with $K$.

Finally, notice that our work opens the path to even more general fitting approaches. Indeed, by discarding the second moment and leveraging only the absolute moments and CFs, one may be able to accurately characterize the distribution of the linear combinations of $t-$RVs with more than one degree of freedom. Moreover, by solely relying on CFs matching with different parameters $r$, one may complete remove the constraint on the number of degrees of freedom and arbitrarily fit the distribution of any linear combination of $t-$RVs to that of a scaled Student's $t$ RV. However, further studies are required on how to achieve this in an optimal and/or simple manner.

\section*{Data Availability}
Codes necessary for reproducing the results of this work are openly available at \url{https://github.com/onel2428/Distribution-of-Sum-of-Students-t-Random-Variables}

\section*{Conflicts of Interest}
No conflicts of interest are identified.

\section*{Funding and Acknowledgment}
This work was supported by the Academy of Finland (6G Flagship Program under Grant 346208) and the Finnish Foundation for Technology Promotion. 

\bibliographystyle{IEEETran}

\bibliography{references}
\end{document}